# Using Monte Carlo simulations to predict the distribution of properties in an ensemble of fluctuating ratchets

Rupsha Mukherjee, Kaustubh Rane[*]

Chemical Engineering, Indian Institute of Technology Gandhinagar, Gujarat, India – 382055

*Corresponding author email id: kaustubhrane@iitgn.ac.in

**Abstract:** We demonstrate the use of Metropolis Monte Carlo simulations and a two-state fluctuating ratchet model to predict the distribution of microscopic properties in a sample of Brownian motors. Our scheme only uses the information about the mean and standard deviation of steady-state velocities of motors. We interpret the results using an analogy between the ensemble of motors and the semi grand canonical ensemble used to study mixtures of chemical species. We observe that increasing the proportion of motors having faster transitions between two states will always increase the mean velocity of the sample, irrespective of the extent of variation in other microscopic properties. On the other hand, changing the proportion of motors based on their potential energies in a particular state has a noticeable effect on mean velocity only when the sample is made homogeneous with respect to most other microscopic properties. The computational approach is easily extendable to models beyond fluctuating ratchets and observables beyond steady-state velocities.

## 1. Introduction

Advances in nanotechnology and synthetic chemistry have facilitated the miniaturization of devices used in several fields. One important breakthrough is the development of molecular or nanoscale motors. Such motors have potential applications in therapeutics and electronics, where they may have to work in large numbers.[1–4] Their successful application requires addressing challenges associated with the scalability of their production, reproducibility of their features, control of their behaviour in large numbers, and physical models suitable for predicting their individual and bulk behaviour. We propose a computational scheme to relate the variation in the features of individual nanoscale motors to their behaviour in large numbers.

The proposed scheme has two applications: 1) Technological: To predict the allowable variation in the microscopic properties of individual motors that results in the desired behaviour in large numbers, and 2) Scientific: To predict the possible variation in the microscopic properties of individual motors from the knowledge of their observed behaviour in large numbers. The technological application is relevant because fabrication of large number of motors having exactly same features may be impractical. The local environment that each motor experiences during an operation may be different, making it difficult to extrapolate the behaviour of ensemble from that of individual motors. The scientific application can be understood from studies performed on the biological motors like Kinesins.[5–7] Kinesins are nanoscale assemblies of proteins that use chemical energy to translate along the structures called microtubules.[8] Their velocities are measured by tracking the motion of several kinesin motors inside an experimental setup. Such a scheme results in a histogram of velocities that indicates a dissimilarity between motors and their local environments inside the setup. The ability to predict the variation in motor-features and local environment that explains the observed velocity-distribution is essential.

Statistical mechanics of mixtures can help relate the properties of an ensemble with those of individual motors. A large collection of motors having different properties or local environments is analogous to a mixture of chemical species. An experimental observable of the collection of motors is akin to a macroscopic property of mixture whose ensemble averages or moments are accessible via experiments. We refer the properties of motors that are difficult to track via experiments as microscopic properties. A shift in the distribution over a particular microscopic property of a motor is analogous to a change in the chemical composition of the mixture. Therefore, predicting the change in the behaviour of large number of motors due to the abovementioned shift is similar to predicting the dependence of macroscopic properties of a mixture on its chemical composition.

The semi-grand canonical (SG) ensemble is used to study mixtures of molecular species. It is defined by constant total number of molecules, volume, temperature, and the chemical potentials of species relative to a reference species.[9–11] The relative chemical potentials govern the compositions of different molecular species. Several studies have used the semi-grand canonical Monte Carlo (SGMC) simulations to compute ensemble-averages of macroscopic properties for mixtures containing finite number of molecular species. Rutledge and co-workers used SGMC simulations as part of a reverse Monte Carlo technique to interpret the experimental measurements for polymeric melts and glasses.[12–14] Their approach took the ensemble-averages of macroscopic properties as input and predicted the distributions over structural features of polymers in a melt. They could explicitly consider the important degrees of freedom of molecules, because their spatial scales were very small. On the other hand, explicitly sampling the molecular degrees of freedom is computationally unfeasible for the ensemble of nanoscale motors, because of their large sizes. They are typically studied via stochastic models that approximate the contributions from molecular-scale motions. The input parameters of such models are the microscopic properties of motors and the output properties are experimental observables. In a SG ensemble of nanoscale motors, the vector of input parameters defines a species. If the model uses $m$ input parameters each of which take $n$ values, then the number of species is $n^m$. Prediction of relative chemical potentials for such a large number of species is impractical.

The SGMC approach of Rutledge is a special case of Maximum Entropy (MaxEnt) methods that are used to gain inference from limited data.[15–17] The method of Lagrange multipliers is used to maximize the informational entropy, with the experimental information serving as constraints. The number of Lagrange multipliers depends on the number of experimental observables. Their estimation may require significant computational expenses and faces technical challenges due to correlations between constraints.[18,19] If a complete probability distribution over an observable is available from experiments on large number of motors, then the MaxEnt approach is not required. The available distribution can be used to perform MC simulations and estimate the distributions over microscopic properties of the system. However, the analogy with SG ensemble can still help in interpreting the results.

Using molecular simulations to compute properties of most artificial and biological nanoscale motors is impractical due to spatial scales involved. Brownian motors form an important subset of nanoscale motors that use Brownian motion and periodic energy-input to perform wide-range of motions.[20–22] One of the simplest models of Brownian motors is the fluctuating-ratchet model that considers a Brownian particle undergoing diffusive motion in the presence of an external field that varies periodically in space and time.[23,24] The resulting motion is modelled via a set of Langevin equations which can be numerically solved to estimate the experimental observables like steady-state velocity, stall-force, etc.[20,21] The input to above models consists of parameters associated with the external field acting on a motor or the potential energy of a motor as a function of space and time, local environment of the motor, and the rate of energy-input via chemical reaction or some other process.[23,24] If the probability distribution over an observable like velocity is available from an experiment on large number of fluctuating ratchets, then MC simulations can be used to estimate the distributions over input parameters. Such a scheme requires the computation of velocity at every MC step by changing the value of at least one input parameter. The application of numerical procedure to solve the Langevin equations at each MC step adds to the computational expenses of the scheme. In the present manuscript, we demonstrate our approach by using the simplest fluctuating ratchet model (two-state ratchet with single particle) and limit our discussion to the steady-state velocity as the experimental observable.

Our MC simulation-based strategy can be used to interpret the experimental results of biological nanoscale motors like kinesins.[6,25] Studies have investigated the effect of specific mutations on the motor-velocities. The experiment involves several kinesin motors and measures the mutation-induced shift in the histogram over velocities. Even if all motors in the sample are chemically equivalent, the distribution over velocities indicate the inhomogeneity in the local environments of the motors.[26–28] The effect of mutations on the velocity of a motor is mediated by its local environment. In order to establish the link between mutations and the observed shift in the velocity-distribution, it is necessary to understand the abovementioned inhomogeneity. The motion of kinesin is too complex to be studied using a simple model like the two-state fluctuating ratchet. The important challenge is the large computational expense involved in coupling a realistic model of kinesin motor with the MC scheme. However, if such challenges are addressed, then the input parameters of the model can be related to the microscopic properties of kinesin motor like the strength of motor-microtubule binding, motor-diffusivity, and the rate of ATP-hydrolysis. All the above properties depend on the chemical nature of constituent proteins and the local environment of kinesin motor. A mutation can shift the velocity-distribution if it changes at least one microscopic property by an amount significantly greater than the variation present in the given sample. Our MC scheme can compute the distributions over input parameters corresponding to the velocity-histograms that are measured before and after mutations. Comparison between the above distributions can help identify input parameters – and microscopic properties – that play important role in the mutation-induced shift in velocities. More specifically, the role of motor-microtubule binding and rate of ATP hydrolysis can be understood.[25] The proposed scheme also enables us to compute the ensemble-averaged microscopic properties that correspond to the mean value of the observed velocities. The information can aid the development of models to be used for kinesin motors.

The manuscript is organized as follows: Section 2 explains the two-state fluctuating ratchet model, the expressions used to calculate motor-velocity, and the MC method. Section 3 details the simulation-parameters, ranges of input-parameters, and the velocity-distributions used to demonstrate our approach. Section 4 discusses the results, and Section 5 concludes the manuscript and provides future-outlook. The Appendix derives relationships between probability distributions obtained from MC simulations and density functions of the SG ensemble.

## 2. Method and Theory

### 2.1 Two-state fluctuating ratchet and calculation of steady-state velocity

The two-state fluctuating ratchet models a Brownian particle capable of translating in a single direction.[20,21,23] Let $r$ denote the position of particle at any given instant, and $D$ be its diffusivity in the absence of external field. We will refer $D$ as bulk diffusivity in rest of the manuscript. The particle experiences an external field that results in the potential energy landscape shown in Figure 1. The saw-tooth-like nature of the potential energy profile is crucial for the performance of motor. The external field switches at regular intervals to change the potential energy profile from I to II and vice versa. Let $k_{12}$ and $k_{21}$ denote the rates of switching from states I to II and II to I, respectively.

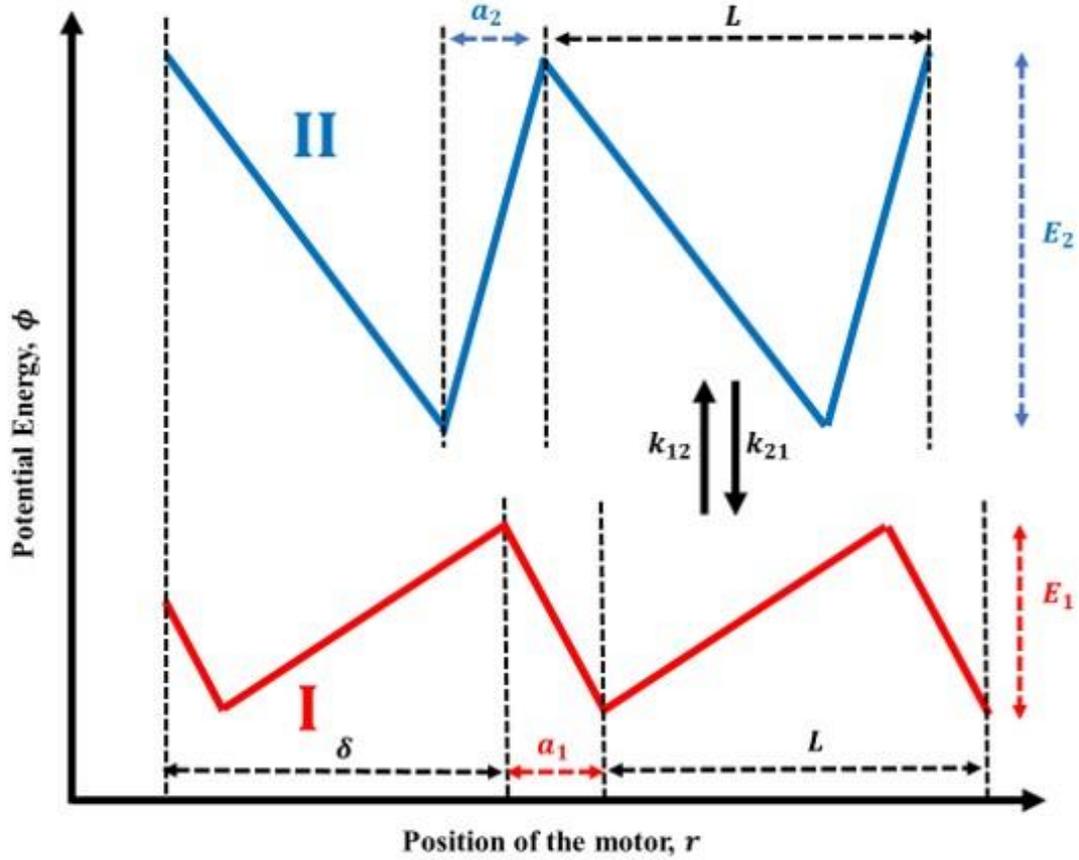

Figure 1. Input parameters of the fluctuating ratchet model. Red and blue lines denote the potential energy profiles of states I and II, respectively. The subscripts 1 and 2 are used to denote parameters of states I and II, respectively. The periodicity $L$ is same for both the states in the present work. The detailed description of parameters is provided in the text.

The input parameters associated with potential energy profiles are also depicted in Figure 1. Let subscript $i$ denote the state of motor. We use $i = 1$ for state I and $i = 2$ for state II. The periodicity of potential energy profile along the direction of motion of the particle is denoted by $L$. We select the same periodicity for states I and II. $E_i$ denotes the difference between minimum and maximum potential energies of state $i$, and $a_i$ denotes the separation between positions having minimum and maximum potential energies. $\delta$ denotes the separation between positions of maximum potential energies of I and II.

The bulk diffusivity $D$ is a function of motor-size and viscosity of the surrounding medium. The rates $k_{12}$ and $k_{21}$ are related to the rates of chemical/physical processes that change the external field acting on the particle. Consequently, $\ln k_{12}$ and $\ln k_{21}$ can be related to the transition state energies of the above processes. $E_i$ indicates the relative stability of motor in state $i$ at two different locations. The parameter $a_i$ is referred as asymmetry parameter. A symmetric saw-tooth profile is obtained when $a_i = L/2$. For a real motor, $L, a_i$ and $\delta$ characterize the variation in local environment due to external field.

In an overdamped environment, the motor-position $r$ is governed by the Langevin equation:[29]

$$\frac{dr}{dt} = -\frac{D}{k_B T}\frac{\partial \phi(r,t)}{\partial r} + \sqrt{2D}\tilde{f}(t) \qquad (1)$$

Here, $k_B$ is the Boltzmann constant, $T$ is the temperature, $\tilde{f}(t)$ is the Gaussian white-noise term, and $\phi(r,t)$ is the potential energy of motor. For the two-state fluctuating ratchet model,

$$\phi(r,t) = \begin{cases} \phi_1(r; E_1, a_1, \delta) & \text{when state I} \\ \phi_2(r; E_2, a_2) & \text{when state II} \end{cases} \qquad (2)$$

Here, $\phi_i$ denotes the potential energy in state $i$. Note that the parameter $\delta$ is included with $\phi_1$ because we use it to denote the shift in potential energy profile of state I relative to that of state II (see Figure 1). Equation (1) can be numerically integrated to compute $r(t)$. The steady-state velocity can be calculated by assuming a linear relationship between $r$ and $t$. The calculation is repeated several times and the average value of velocity is used for further analysis. The above procedure can be used in principle to calculate velocity in MC simulations. However, performing the procedure at every MC step increases the computational expenses. The alternative is to consider the Fokker-Planck equation[20,29] that describes the evolution of probability density $\rho_i(r,t)$ of finding the motor in state $i$ at position $r$ and time $t$:

$$\frac{\partial \rho_1}{\partial t} = D\frac{\partial}{\partial r}\left(\frac{1}{k_B T}\frac{\partial \phi_1}{\partial r}\rho_1 + \frac{\partial}{\partial r}\rho_1\right) - k_{12}\rho_1 + k_{21}\rho_2 \qquad (3)$$

$$\frac{\partial \rho_2}{\partial t} = D\frac{\partial}{\partial r}\left(\frac{1}{k_B T}\frac{\partial \phi_2}{\partial r}\rho_2 + \frac{\partial}{\partial r}\rho_2\right) - k_{21}\rho_2 + k_{12}\rho_1 \qquad (4)$$

Wang et al. proposed an algorithm to solve the above equations by considering a discrete Markov process along $r$ and deriving the transition probabilities by using local approximate solutions.[29] The algorithm calculates steady-state velocity as a solution of linear equations, and can be straightforwardly implemented via matrix operations in the programming language of choice. We use the above method to calculate velocities because it is less expensive than numerically solving Equation (1) at every MC step. The following expressions are used to calculate velocity:

$$v = L \sum_{n=1}^{N_L} [(\boldsymbol{T}_+ - \boldsymbol{T}_-)\boldsymbol{p}^s]_n \tag{5}$$

Here, $N_L$ denotes the number of discrete points considered along $r$. $\boldsymbol{p}^s$ is defined as

$$\boldsymbol{p}^s = \begin{bmatrix} p_1^s(r_1) \\ . \\ . \\ p_1^s(r_{N_L}) \\ p_2^s(r_1) \\ . \\ . \\ p_2^s(r_{N_L}) \end{bmatrix} \tag{6}$$

Where, $p_i^s(r_n)$ denotes the steady-state probability of finding the motor in state $i$ at $r_n$. $\boldsymbol{p}^s$ is evaluated by solving the following linear equation:

$$\boldsymbol{M}\boldsymbol{p}^s = 0 \tag{7}$$

Subjected to the normalization constraints $\sum_{n=1}^{N_L}[p_1^s(r_n) + p_2^s(r_n)] = 1$. $\boldsymbol{M}, \boldsymbol{T}_+$ and $\boldsymbol{T}_-$ are $2N_L \times 2N_L$ matrices containing the probabilities for transitioning from one discrete state to the adjacent state. $\boldsymbol{M}$ is given by,

$$\boldsymbol{M} = \begin{bmatrix} \boldsymbol{T}^{(1)} - \boldsymbol{K}_{12} & \boldsymbol{K}_{21} \\ \boldsymbol{K}_{12} & \boldsymbol{T}^{(2)} - \boldsymbol{K}_{21} \end{bmatrix} \tag{8}$$

Here, $\boldsymbol{K}_{12}$ and $\boldsymbol{K}_{21}$ are the $N_L \times N_L$ diagonal matrices containing $k_{12}$ and $k_{21}$ as elements, respectively. $\boldsymbol{T}^{(i)}$ ($i$ = 1 and 2 for states I and II, respectively) are $N_L \times N_L$ tridiagonal matrices having the following elements:

$$T_{n,n}^{(i)} = -\left(F_{n+1/2}^{(i)} + B_{n-1/2}^{(i)}\right) \tag{9}$$

$$T_{n-1,n}^{(i)} = B_{n-1/2}^{(i)} \tag{10}$$

$$T_{n+1,n}^{(i)} = F_{n+1/2}^{(i)} \tag{11}$$

$$T_{N_L,1}^{(i)} = B_{1/2}^{(i)} \tag{12}$$

$$T_{1,N_L}^{(i)} = F_{N_L+1/2}^{(i)} \tag{13}$$

$F_{n+1/2}^{(i)}$ denotes the rate of transitioning in the forward direction from $r_n$ to $r_{n+1}$. $B_{n-1/2}^{(i)}$ denotes the rate of transitioning in the backward direction from $r_n$ to $r_{n-1}$. They are evaluated as follows:

$$F_{n+1/2}^{(i)} = \frac{D}{(\Delta r)^2} \frac{\Delta\phi_{n+1/2}^{(i)}/k_B T}{\exp\left(\Delta\phi_{n+1/2}^{(i)}/k_B T\right) - 1} \tag{14}$$

$$B_{n-1/2}^{(i)} = \frac{D}{(\Delta r)^2} \frac{-\Delta\phi_{n-1/2}^{(i)}/k_B T}{\exp\left(-\Delta\phi_{n-1/2}^{(i)}/k_B T\right) - 1} \tag{15}$$

Here, $\Delta\phi_{n+1/2}^{(i)} = \phi_i(r_{n+1}) - \phi_i(r_n)$ and $\Delta\phi_{n-1/2}^{(i)} = \phi_i(r_n) - \phi_i(r_{n-1})$. Finally, $\boldsymbol{T}_+$ and $\boldsymbol{T}_-$ are $2N_L \times 2N_L$ matrices whose all elements are zero except,

$$(T_+)_{1,N_L} = F_{N_L+1/2}^{(1)} \tag{16}$$

$$(T_+)_{N_L+1,2N_L} = F_{N_L+1/2}^{(2)} \tag{17}$$

$$(T_-)_{N_L,1} = B_{1/2}^{(1)} \tag{18}$$

$$(T_-)_{2N_L,N_L+1} = B_{1/2}^{(2)} \tag{19}$$

### 2.2 Monte Carlo simulations

Let $X$ denote the experimental observable and $\pi(X)$ denote the probability density of observing $X$. Then, the probability of observing $X$ in the interval $[X, X + dX]$ is given by $\Pi(X) = \pi(X)dX$. In the present manuscript, $X \equiv v$ is the steady-state velocity of motor, and $\Pi(v)$ is supposedly derived from the experimental measurements on several motors. We assume that the above distribution is due to the intrinsic heterogeneity of motors and the variation of local environment. The uncertainty in experimental measurements is assumed to have negligible contribution. More information about the experimental methods and real experimental data are required to account for the experimental uncertainties. Such an analysis is outside the scope of the present work, because we demonstrate our approach using a model system. Let $\boldsymbol{x}$ denote the vector of $m$ parameters or microscopic properties input to the model. Let $p(\boldsymbol{x})$ denote the probability density of observing $\boldsymbol{x}$ in the given sample of motors. We assume that the sample has large number of motors and the components of $\boldsymbol{x}$ can be approximated as continuous variables. Therefore, the probability of observing $\boldsymbol{x}$ in the interval $[\boldsymbol{x}, \boldsymbol{x} + d\boldsymbol{x}]$ is given by $P(\boldsymbol{x}) = p(\boldsymbol{x})dx_1 \ldots dx_m$. In the present manuscript, $x_1, \ldots, x_m$ are input parameters to the two-state fluctuating ratchet model. If $v(\boldsymbol{x})$ denotes the output velocity, then

$$\Pi(X) = \int \ldots \int \delta(X - v(\boldsymbol{x}))p(\boldsymbol{x})dx_1 \ldots dx_m \tag{20}$$

$$\int \ldots \int p(\boldsymbol{x})dx_1 \ldots dx_m = 1 \tag{21}$$

Here, $\delta$ denotes the Dirac delta function. The limits of integrals depend on the possible values of parameters. The probability distributions over selected parameters are related to $p(\boldsymbol{x})$ as follows:

$$\rho^l(x_1, \ldots x_l) = \int \ldots \int p(\boldsymbol{x})dx_{l+1} \ldots dx_m \tag{22}$$

Here, $\rho^l$ denotes the probability density of observing $x_1, \ldots, x_l$. We discretize $x_1, \ldots, x_m$ to give $\aleph$ microstates. Let $\boldsymbol{x}^{(i)}$ denote the vector of input parameters corresponding to $i^{th}$ microstate. Then, the probability $p_d(\boldsymbol{x}^{(i)})$ of observing $i^{th}$ microstate is given by

$$p_d(\mathbf{x}^{(i)}) = \frac{\Pi(v^{(i)})}{Q} \quad (23)$$

$$Q = \sum_{i=1}^{\aleph} \Pi(v^{(i)}) \quad (24)$$

Equation (23) provides the probability of observing a particular combination of input parameters $x_1, \ldots, x_m$ that results in a given value of observable. However, the probability distributions over selected parameters are generally of interest. Following Equation (22), the discrete versions of such distributions are given by,

$$\rho_d^l(x_1, \ldots, x_l) = \sum_{i=1}^{\aleph_{m-l}} p_d(\mathbf{x}^{(i)}) \quad (25)$$

Here, $\aleph_{m-l}$ denotes the number of microstates in which $x_1, \ldots, x_l$ are fixed. Note that it is impractical to compute $\rho_d^l$ by using Equations (23) and (25) because of several possible combinations of values taken by $x_{l+1}, \ldots, x_m$. We therefore use a Markov chain Monte Carlo scheme involving the following steps: 1) Starting with suitable values of $x_1, \ldots, x_m$. 2) Selecting a parameter $x_i$ with uniform probability. 3) Proposing a change in the microstate by changing the value of selected parameter by $\Delta x_i$ or $-\Delta x_i$ with equal probability. $\Delta x_i$ is the discretization used for $x_i$. 4) Accepting the proposed change via Metropolis-Hastings acceptance criterion:

$$pacc^{(i \to j)} = \min\left[1, \frac{p_d^{(j)}}{p_d^{(i)}}\right] \quad (26)$$

Here, $pacc^{(i \to j)}$ denotes the probability of accepting the move from microstate $i$ to microstate $j$ and $p_d^{(i)}$ denotes the probability of observing $i^{th}$ microstate as given by Equation (23). 5) Updating the histograms over selected parameters. 6) Computing the probability distributions $\rho_d^l$ over input parameters and ensemble-averages from the gathered histograms. The Monte Carlo simulations described above can be considered equivalent to the simulations performed in a semi grand canonical (SG) ensemble. We compare the two approaches in the Appendix.

## 3. Simulation details

We use a Gaussian distribution over velocity to demonstrate our approach. Following the description in Section 2.2,

$$\Pi(v) = \frac{1}{\sigma_v\sqrt{2\pi}} \exp\left[-\frac{1}{2}\left(\frac{v-\mu_v}{\sigma_v}\right)^2\right] \tag{33}$$

Here, $\mu_v$ and $\sigma_v$ denote the mean and standard deviation, respectively. In future applications, such a distribution may be obtained from experiments and may deviate from Gaussian distribution. Five distributions having different mean velocities, but same standard deviation are selected. We use $\mu_v = 0.4, 0.45, 0.5, 0.55,$ and $0.6 \mu m/s$ for five distributions. The standard deviation is $0.035 \mu m/s$. The mean velocities and standard deviation are selected to be of similar order of magnitude as those observed in experiments with Kinesin motors.[26–28]

The fluctuating ratchet model has eight input parameters: $E_1, E_2, k_{12} = k_{21} = k, D, a_1, a_2, \delta$, and $L$. The description of above parameters is provided in Section 2.1. We assume that the rate constants for switching from states I to II and II to I are same. We work with the logarithm of rate constant ($\ln k$) to handle the range of $k$ spanning orders of magnitude. $L$ is constant in all our calculations. We discretize the ranges to perform Monte Carlo (MC) simulations. Table 1 provides the ranges of all parameters and intervals used for discretization. All energies are expressed in the units of $k_B T$, where $k_B$ is the Boltzmann constant and $T$ is the absolute temperature. Note that the parameters are restricted to values typically observed or used in the studies performed with kinesin motors. We also note that the ranges in Table 1 are selected to limit the parameter-space to be sampled in MC simulations, while ensuring that the values are comparable to those observed for realistic nanoscale motors like kinesins. That being said, the two-state fluctuating ratchet model is not a good descriptor of the kinesin motion and the results should not be interpreted as those of kinesin motors.

*Table 1. Ranges of the input parameters of the fluctuating ratchet model*

| Parameter | Range | Interval used for discretization |
|---|---|---|
| $E_1$ | $-8$ to $2\ k_BT$ | $1\ k_BT$ |
| $E_2$ | $11.0$ to $48.0\ k_BT$ | $1\ k_BT$ |
| $\ln k$ | $3.0$ to $7.0\ k_BT$ | $0.1\ k_BT$ |
| $D$ | $1500$ to $6250\ nm^2/s$ | $50\ nm^2/s$ |
| $a_1$ | $1.0$ to $8.0\ nm$ | $0.1\ nm$ |
| $a_2$ | $1.0$ to $8.0\ nm$ | $0.1\ nm$ |
| $\delta$ | $1.0$ to $8.0\ nm$ | $0.1\ nm$ |
| $L$ | $8.0\ nm$ | $0.08\ nm$ |

The parameter ranges shown in Table 1 impose a non-natural constraint on the estimated distribution of microscopic properties. However, such constraints are necessary for studying real motors whose microscopic properties are restricted by nature or design. For example, in real motors, the ranges of $E_1$ and $E_2$ may never overlap, or the value of $D$ may be limited by the size of motor or the nature of surrounding medium. In such cases, sampling an unrestricted parameter space may lead to unrealistic estimates of the distribution over microscopic features.

The steady-state velocity is computed from the knowledge of $2N_L \times 2N_L$ sized transition probability matrices $\boldsymbol{T}_+$ and $\boldsymbol{T}_-$, and $2N_L$ sized probability vector $\boldsymbol{p}^s$ using Equation (5). The elements of $\boldsymbol{T}_+$ and $\boldsymbol{T}_-$ are computed using Equations (14) - (16). The elements of $\boldsymbol{p}^s$ are obtained by solving Equation (7), where $\boldsymbol{M}$ is computed using Equations (8) – (13). We use the interval of $0.08\ nm$ to discretize $x$. That is, $\Delta x$ in Equations (14) and (15) is $0.08\ nm$. Since $x$ ranges from 0 to $L = 8.0\ nm$, the number of points $N_L$ is 100.

All MC simulations are started with the following values of input parameters to the fluctuating ratchet model $E_1 = -6.0\, k_B T, E_2 = 25.0\, k_B T, \ln k = 5.87\, k_B T, D = 5330\, \frac{nm^2}{s}, a_1 = 6.4\, nm, a_2 = 6.4\, nm,$ and $\delta = 8.0\, nm$. The above parameters result in $v = 0.5 \mu m/s$. A change in any one of the above parameters is proposed with equal probability and accepted using Metropolis-Hastings acceptance criterion. The criterion provided in Equation (26) is used with $p_d^{(i)} \propto \Pi(v^{(i)})$. $\Pi(v^{(i)})$ is computed using Equation (33) where $v^{(i)}$ denotes the velocity corresponding to the $i^{th}$ microstate. The following three simulations are performed for each velocity distribution:

MC1: We only allow $E_1$, $E_2$ and $\ln k$ to vary during the simulations. This is same as assuming that the sample of fluctuating ratchets have same bulk diffusivities, and locations of teeth. However, the heights of teeth in states I and II may differ, as also the rate of transition between states. Figure 2a shows the allowed variations in MC1.

MC2: We keep $D$ constant and allow the variation in $E_1, E_2, \ln k, a_1, a_2,$ and $\delta$. This is same as assuming that the sample of motors have same bulk diffusivities. Figure 2b shows the allowed variations in MC2.

MC3: All seven parameters $(E_1, E_2, \ln k, D, a_1, a_2, \delta)$ are allowed to vary.

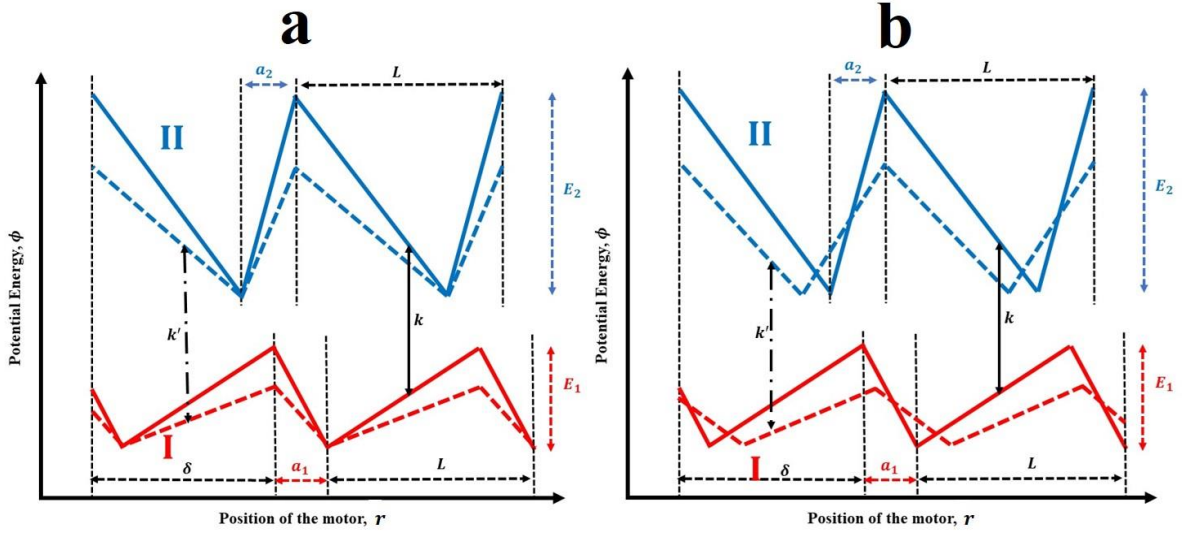

Figure 2. Variations allowed in the input parameters of the fluctuating ratchet model. Red and blue lines denote the potential energy profiles of states I and II, respectively. a) The dashed red and blue lines denote the variation in potential energy profiles that is allowed in MC1 simulations. b) The dashed red and blue lines denote the variation in potential energy profiles that is allowed in MC2 and MC3 simulations.

We compute the one variable distributions $\rho_d^1(x_i)$ and two variable distributions $\rho_d^2(x_i, x_j)$ from the above simulations by collecting the one- and two-dimensional histograms, respectively. Here, $x_i$ denotes an input parameter that is allowed to vary during the simulations. We calculate $\rho_d^2(x_i, x_j)$ only when $x_i, x_j$ are one of $E_1, E_2$, and $\ln k$.

All MC simulations are performed for $10^6$ MC steps, and we use the data from last $8 \times 10^5$ steps to compute $\rho_d^1$ and $\rho_d^2$. Each MC simulation is performed four times to predict the calculation-errors in the above properties. Python programming language was used to perform all calculations.

## 4. Results and Discussions:

### 4.1 Monte Carlo sampling

A Gaussian distribution over velocity is provided as input to our MC scheme. We also calculate velocity-distributions by gathering histograms during simulations. Figure 3 compares the distributions calculated from MC simulations with the input Gaussian distributions. Statistical uncertainties are estimated from four independent simulation runs. We observe good compliance between the input distributions and those computed from MC1, MC2, and MC3 simulations. Note that the velocity-distributions shown in Figure 3 and all results discussed in rest of the manuscript are calculated from data generated during last 800,000 MC steps.

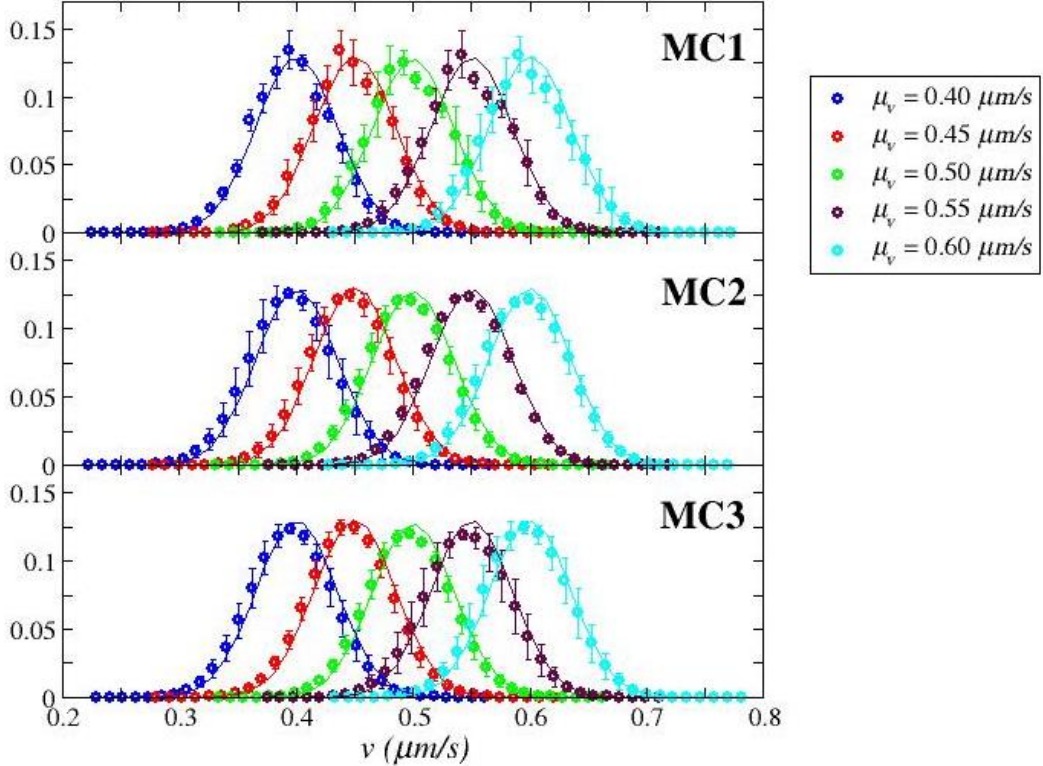

Figure 3. Comparison between input velocity distributions and those generated from the Monte Carlo simulations. Top to bottom rows show results obtained from MC1, MC2, and MC3 simulations, respectively. Blue, red, green, maroon, and cyan symbols denote velocity distributions having average velocities of 0.4, 0.45, 0.50, 0.55, and 0.6 $\mu m/s$, respectively. Straight lines show the Gaussian distributions (Equation (33)) that are used as input to the Monte Carlo simulations. Circles denote the velocity-distributions computed from the Monte Carlo simulations.

Each MC step involves selecting an input parameter, proposing its change, and accepting or rejecting the change via Metropolis-Hastings acceptance criterion. Table 2 provides the percentage acceptance of changes proposed in each parameter for Monte Carlo simulations. The acceptance is similar for MC1, MC2, and MC3 simulations. We observe that the intervals selected for discretization allow satisfactory transitions. This is also seen from Figure 4 which plots the variation of input parameters with MC steps for the simulations performed with input velocity-distribution having mean value of 0.5 $\mu m/s$. We have only shown the results from 200000$^{th}$ to 300000$^{th}$ MC step in Figure 4.

*Table 2. Percentage-acceptance of changes proposed in input-parameters during the MC simulations*

| Parameter changed | Percentage of proposed changes accepted |
|---|---|
| $E_1$ | 95 |
| $E_2$ | 99 |
| $\ln k$ | 95 |
| $D$ | 100 |
| $a_1$ | 97 |
| $a_2$ | 95 |
| $\delta$ | 95 |

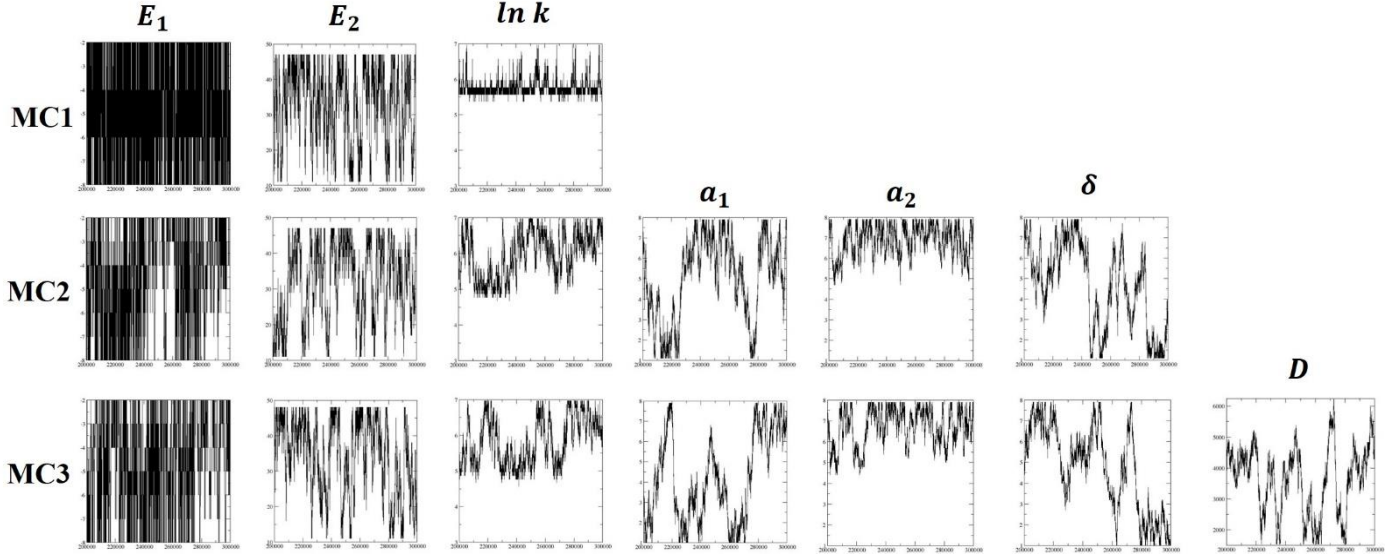

Figure 4. Variation of input parameters with Monte Carlo steps. Results are shown for Monte Carlo simulations performed with input velocity-distribution having mean value of 0.5 $\mu m/s$. All properties are plotted vs the number of Monte Carlo steps. Rows from top to bottom denote results obtained from MC1, MC2, and MC3 simulations, respectively. Columns from left to right show the variation of $E_1, E_2, \ln k, a_1, a_2, \delta$, and $D$, respectively. $E_1$, $E_2$, and $\ln k$ are plotted in the units of $k_B T$. $a_1, a_2$, and $\delta$ are shown in $nm$. $D$ is shown in $nm^2/s$. Results are plotted between 200000$^{th}$ and 300000$^{th}$ MC steps.

**4.2 MC1 simulations (When only $E_1, E_2$, and $\ln k$ are allowed to vary)**

Figure 5 plots the one-variable distributions of $E_1, E_2$, and $\ln k$. Following the analogy with semi-grand canonical (SG) ensemble in the Appendix, $\rho_d^1(E_1), \rho_d^1(E_2)$, and $\rho_d^1(\ln k)$ denote the proportion of motors having particular values of $E_1, E_2$, and $\ln k$, respectively.

Figure 5a shows that all profiles of $\rho_d^1(E_1)$ show a minimum at $E_1 \approx -6.7 \ k_B T$. The profiles become flatter with decrease in mean velocity $\mu_v$ of the input Gaussian distribution. That is, increasing the proportion of motors having larger values of $E_1$ (smaller magnitude) increases the average velocity of the sample. Recollect that the magnitude of $E_1$ denotes the difference between maximum and minimum potential energy of motor in State I. Therefore, increasing the proportion of motors having flatter potential energy profile in state I increases the average velocity of the sample.

Figure 5b shows that the profiles of $\rho_d^1(E_2)$ are statistically similar for most of the range. Therefore, changing the composition of motors based on their potential energies in state II has negligible effect on the mean velocity of the sample.

Figure 5c shows that all profiles of $\rho_d^1(\ln k)$ show a distinct peak, which indicates that the velocity is more sensitive to change in $\ln k$ than $E_1$ or $E_2$. The peak shifts to higher values of $\ln k$ with increasing $\mu_v$, or increasing the proportion of motors having greater $\ln k$ increases the average velocity of the sample. If the transitions between states $I$ and $II$ occur via chemical reactions, then $\ln k$ is proportional to the negative of activation energy $-E_A$ (in the units of $k_B T$) along a suitable reaction-coordinate. Since $E_A$ denotes the potential energy of transition-state relative to state I or II, Figure 5c implies that increasing the proportion of motors having more stable transition states increases the average velocity of the sample.

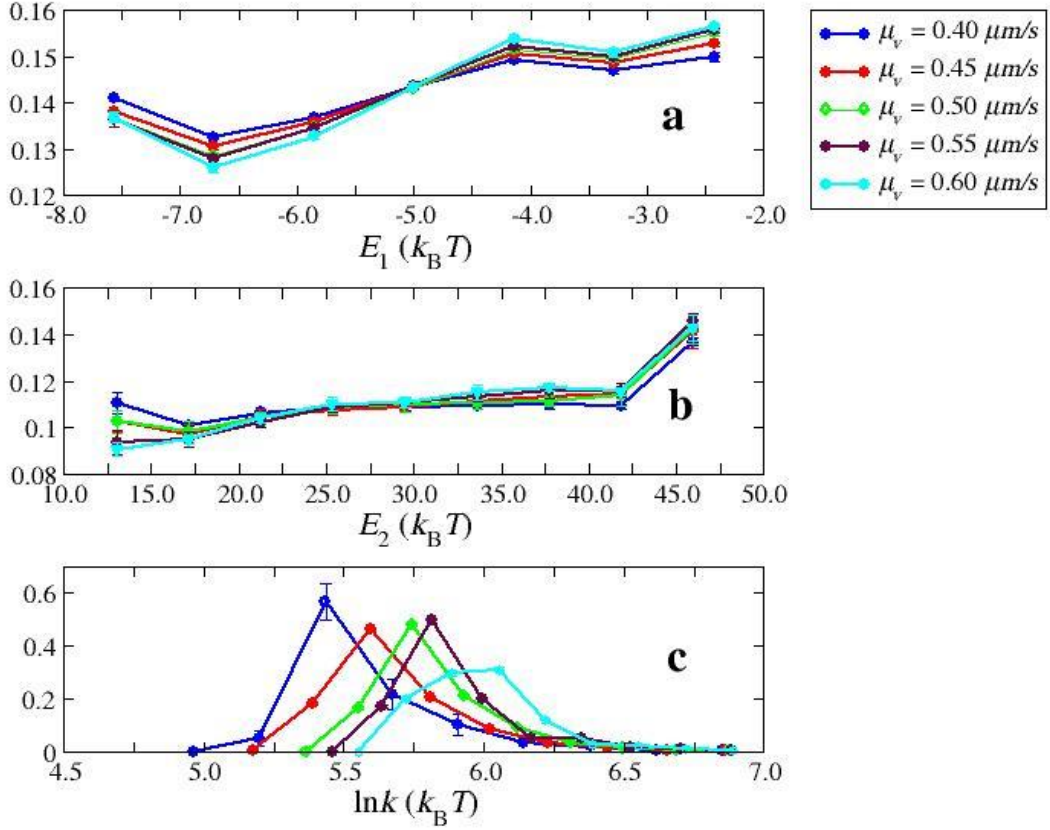

Figure 5. One-variable distributions of $E_1$, $E_2$, and $\ln k$ generated from MC1 simulations. Figures a, b, and c show distributions of $E_1$, $E_2$, and $\ln k$, respectively. Blue, red, green, maroon, and cyan symbols denote results corresponding to the input velocity-distributions having average velocities of 0.4, 0.45, 0.50, 0.55, and 0.6 $\mu m/s$, respectively.

We now discuss the effect of changing the composition of motors having same values of two parameters. Figure 6 shows the two-variable distributions $\rho_d^2(E_1, E_2), \rho_d^2(E_1, \ln k)$, and $\rho_d^2(E_2, \ln k)$. Following the analogy with SG ensemble in the Appendix, they indicate the proportion of motors having particular values of mentioned parameters.

Figures 6a-6e show that $\rho_d^2(E_1, E_2)$ has negligible variation over the sampled parameter-space when compared to $\rho_d^2(\ln k, E_1)$ and $\rho_d^2(\ln k, E_2)$. This shows that altering the proportion of motors having particular values of $E_1$ and $E_2$ has negligible effect on the average velocity of sample when $\ln k$ is allowed to vary. Figures 6a-6e can be also interpreted as follows: If the given sample of motors shows a Gaussian distribution over velocities, there exists negligible

correlation between $E_1$ and $E_2$. That is, the potential energies of motors in states I and II have negligible correlation.

Figures 6f-6j plot $\rho_d^2(\ln k, E_1)$ for different values of $\mu_v$. We observe that $\rho_d^2(\ln k, E_1) \to 0$ for several combinations of $E_1$ and $\ln k$. This implies that the sample lacks motors having those values of $E_1$ and $\ln k$. If the goal is fabrication of motors having a particular $\mu_v$, then the above combinations of $E_1$ and $\ln k$ must be avoided. We notice that the proportion of motors having intermediate values of $\ln k$ and smaller values (larger magnitudes) of $E_1$ is greater than the others. Also, the above $\ln k - E_1$ region slightly shifts towards greater values of $\ln k$ with increasing $\mu_v$.

$\rho_d^2(\ln k, E_2)$ shows stronger dependence on $\mu_v$ than $\rho_d^2(\ln k, E_1)$ as seen in Figures 6k-6o. The distribution shifts towards larger values of $\ln k$ with increasing $\mu_v$. Here, $\rho_d^2(\ln k, E_2) \to 0$ over a larger parameter-space. The proportion of motors having high or low values of both $E_2$ and $\ln k$ is negligible. Therefore, the sample lacks motors having extreme values of both $E_2$ and $\ln k$. Alternatively, extreme values of both $E_2$ and $\ln k$ must be avoided to generate a sample of motors having Gaussian velocity-distribution.

The observed shifts in $\rho_d^2(\ln k, E_1)$ and $\rho_d^2(\ln k, E_2)$ with $\mu_v$ have technological relevance. Chemical modifications at molecular scales may be used to alter parameters like $E_1$, $E_2$ and $\ln k$ of real nanoscale motors. An important example is the replacement of certain amino acid residues (mutations) of biological motors to alter their microscopic properties, and thereby, their performance. Such modifications may affect the potential energies of motors in states I and/or II, as well as the rates of transitions between two states. If the transitions occur via chemical reactions, then the chemical modifications affecting potential energies in states I and II may also affect the transition states. It is then useful to identify the path in $E_1 - \ln k$ or $E_2 - \ln k$ space that can result in the desired shift in mean velocity $\mu_v$ of the sample. For example,

Figures 6k-6o show that modifications increasing both $E_2$ and $\ln k$ will increase $\mu_v$, as the distribution shifts towards top right corner with increase in $\mu_v$.

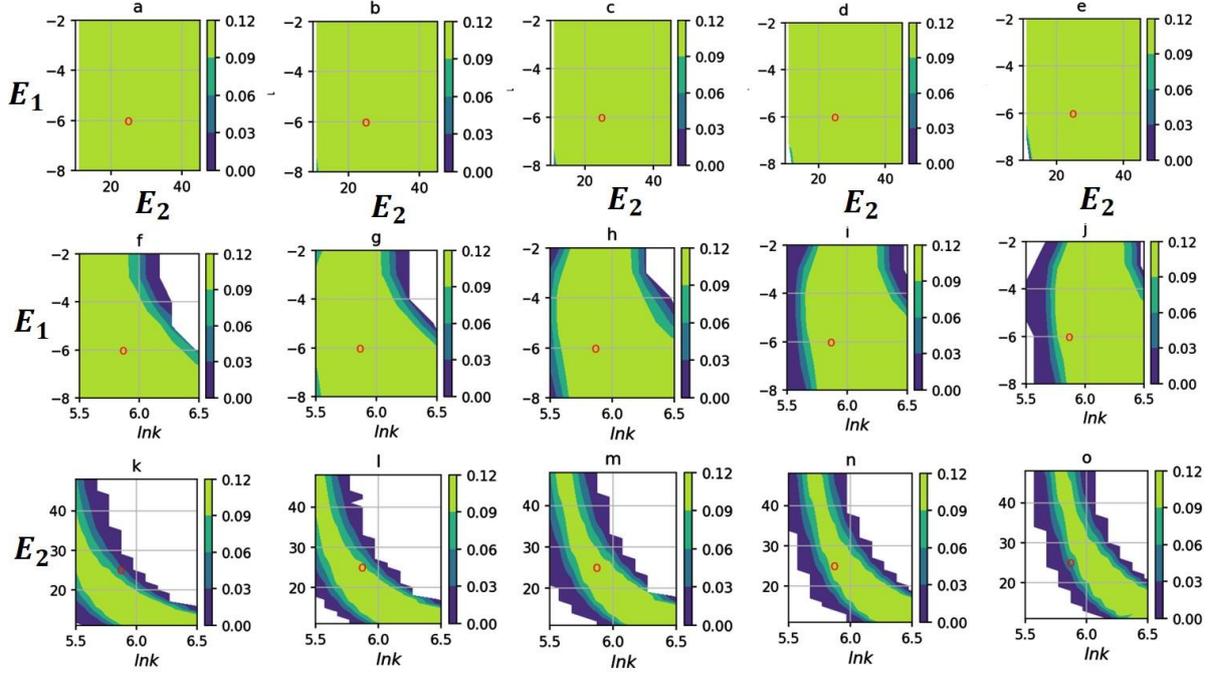

Figure 6. Two-variable distributions obtained from MC1 simulations. Figures a to e show the joint distributions of $E_1$ and $E_2$. Figures f to j show the joint distributions of $E_1$ and $\ln k$. Figures k to o show the joint distributions of $E_2$ and $\ln k$. The figures from left to right in each row denote results corresponding to the input velocity-distributions having average velocities of 0.4, 0.45, 0.50, 0.55, and 0.6 $\mu m/s$, respectively. The red circle in each figure corresponds to $E_1 = -6.0\ k_B T, E_2 = 25.0\ k_B T$, and $\ln k = 5.87\ k_B T$.

**4.2 MC2 simulations (When all parameters except $D$ are allowed to vary)**

Figure 7 plots one-variable distributions of $E_1, E_2, \ln k, a_1, a_2$, and $\delta$. Following the analogy with semi-grand canonical (SG) ensemble in the Appendix, $\rho_d^1(E_1), \rho_d^1(E_2), \rho_d^1(\ln k), \rho_d^1(a_1), \rho_d^1(a_2)$, and $\rho_d^1(\delta)$ denote the proportion of motors having particular values of $E_1, E_2, \ln k, a_1, a_2$, and $\delta$, respectively.

Figures 7a-7e show that $\rho_d^1(E_1), \rho_d^1(E_2), \rho_d^1(a_1)$, and $\rho_d^1(a_2)$ change monotonously, whereas $\rho_d^1(\delta)$ has a minimum. All of them show weak dependence on $\mu_v$. $E_1, E_2, a_1, a_2$, and $\delta$ are associated with the potential energy profiles of states I and II. Therefore, chemical modifications only altering the potential energy profiles of states I and II have negligible effect on the mean velocity of sample.

Figure 7f shows that $\rho_d^1(\ln k)$ shows a distinct peak that shifts to higher values of $\ln k$ with increasing $\mu_v$. That is, increasing the proportion of motors having greater $\ln k$, or more stable transition state between I and II, increases the average velocity of the sample.

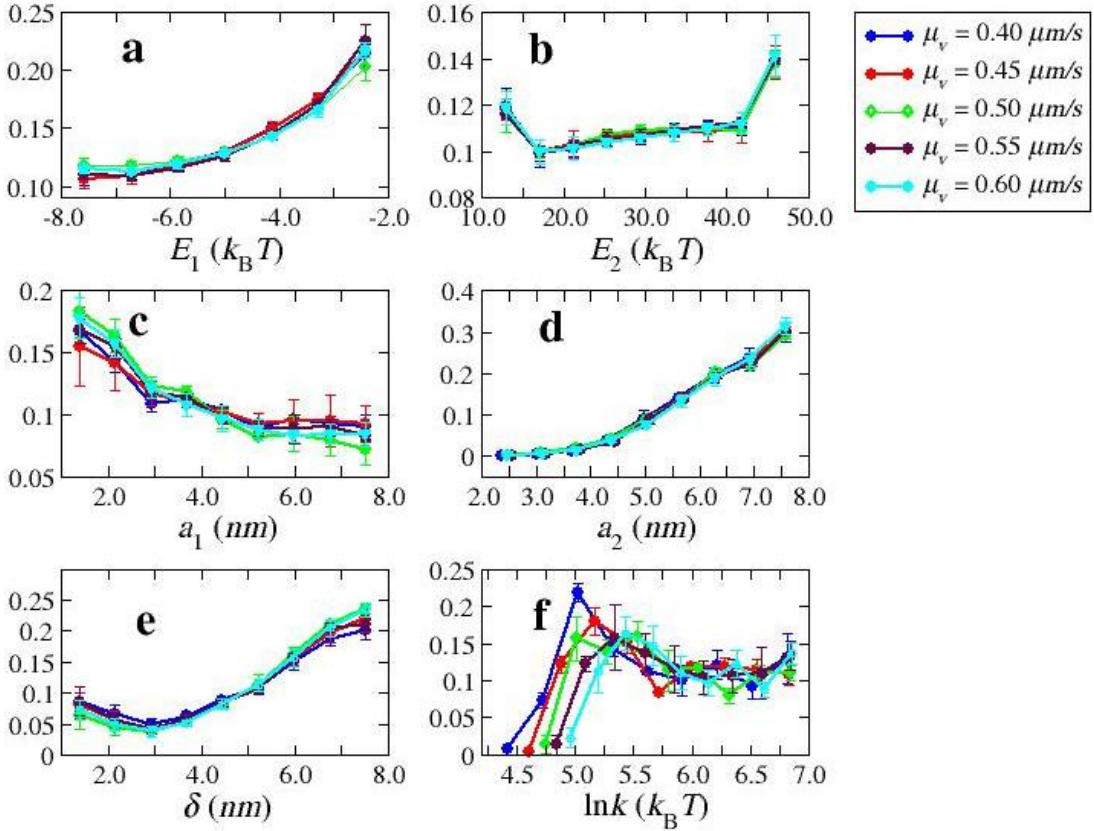

Figure 7. One-variable distributions of $E_1$, $E_2$, $a_1$, $a_2$, $\delta$, and $\ln k$ generated from MC2 simulations. Figures a, b, c, d, e, and f show distributions of $E_1$, $E_2$, $a_1$, $a_2$, $\delta$, and $\ln k$, respectively. Blue, red, green, maroon, and cyan symbols denote results corresponding to the input velocity-distributions having average velocities of 0.4, 0.45, 0.50, 0.55, and 0.6 $\mu m/s$, respectively.

Figure 8 shows the two-variable distributions $\rho_d^2(E_1, E_2), \rho_d^2(E_1, \ln k)$, and $\rho_d^2(E_2, \ln k)$. We observe that $\rho_d^2(E_1, E_2)$ has negligible variation over the sampled parameter-space. Figures 8f-8o show that $\rho_d^2(\ln k, E_1)$ and $\rho_d^2(\ln k, E_2)$ are negligible for smaller values of $\ln k$. We also observe that the distribution shifts towards greater values of $\ln k$ with increasing $\mu_v$. The shift is greater than that seen for MC1 in Figure 6. Negligible variation is observed along $E_1$ or $E_2$. Thus, increasing the proportion of motors having larger $\ln k$ (more stable transition states) is the only way to increase the average velocity of sample that is polydisperse with respect to $E_1, E_2, \ln k, a_1, a_2,$ and $\delta$.

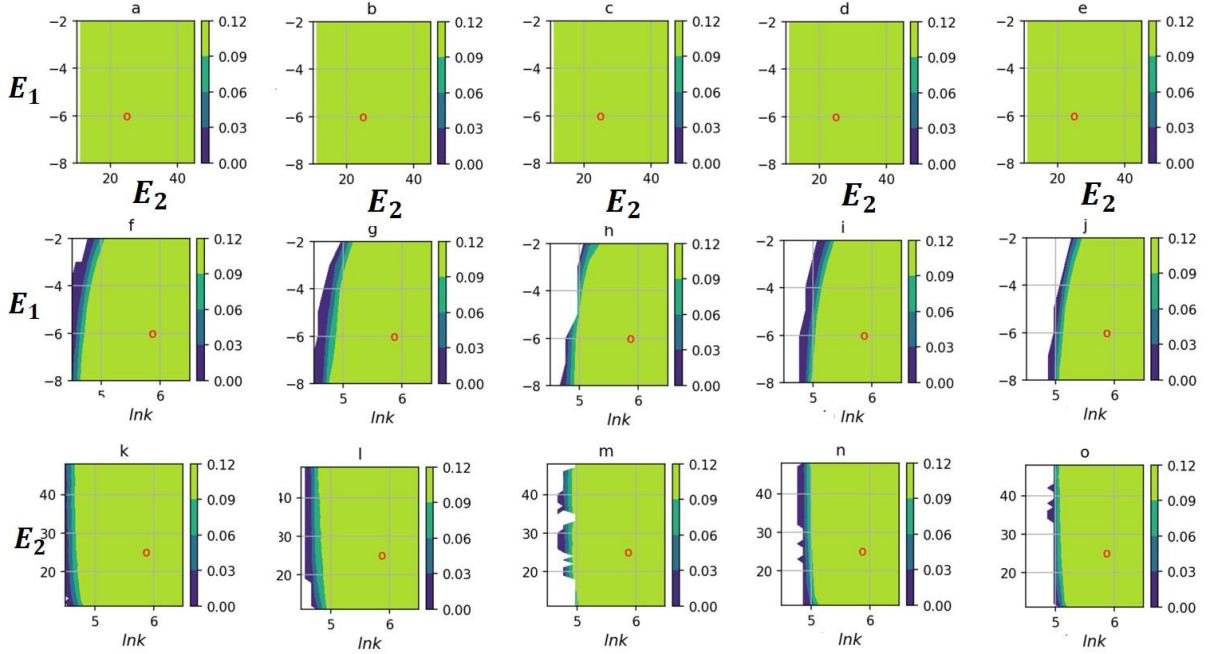

Figure 8. Two-variable distributions obtained from MC2 simulations. Figures a to e show the joint distributions of $E_1$ and $E_2$. Figures f to j show the joint distributions of $E_1$ and $\ln k$. Figures k to o show the joint distributions of $E_2$ and $\ln k$. The figures from left to right in each row denote results corresponding to the input velocity-distributions having average velocities of 0.4, 0.45, 0.50, 0.55, and 0.6 $\mu m/s$, respectively. The red circle in each figure corresponds to $E_1 = -6.0 \ k_B T, E_2 = 25.0 \ k_B T,$ and $\ln k = 5.87 \ k_B T$.

**4.3 MC3 simulations (When all parameters are allowed to vary)**

Figure 9 plots the one-variable distributions of $E_1, E_2, a_1, a_2, \delta, \ln k$ and $D$. Following the analogy with semi-grand canonical (SG) ensemble in the Appendix, $\rho_d^1(E_1), \rho_d^1(E_2), \rho_d^1(a_1), \rho_d^1(a_2), \rho_d^1(\delta), \rho_d^1(\ln k),$ and $\rho_d^1(D)$, denote the proportion of motors having the particular values of $E_1, E_2, a_1, a_2, \delta, \ln k,$ and $D$, respectively. We observe that the profiles of $\rho_d^1(E_1), \rho_d^1(E_2), \rho_d^1(a_1), \rho_d^1(a_2), \rho_d^1(\delta),$ and $\rho_d^1(\ln k)$ are similar to those of MC2.

$\rho_d^1(D)$ is statistically same for the sampled range, and shows negligible dependence on $\mu_v$. Figure 10 plots the two-variable distributions $\rho_d^2(E_1, E_2), \rho_d^2(E_1, \ln k)$, and $\rho_d^2(E_2, \ln k)$. We see that the two-variable distributions are also similar to those observed for MC2.

The above observations indicate that allowing the variation of $D$ does not significantly affect the distribution of parameters. The bulk diffusivity $D$ depends on the size of motor via the Stokes-Einstein relation. Therefore, polydispersity with respect to size may not affect the ability to change the average velocity of sample by altering the proportion of motors based on $E_1$ or $\ln k$ values.

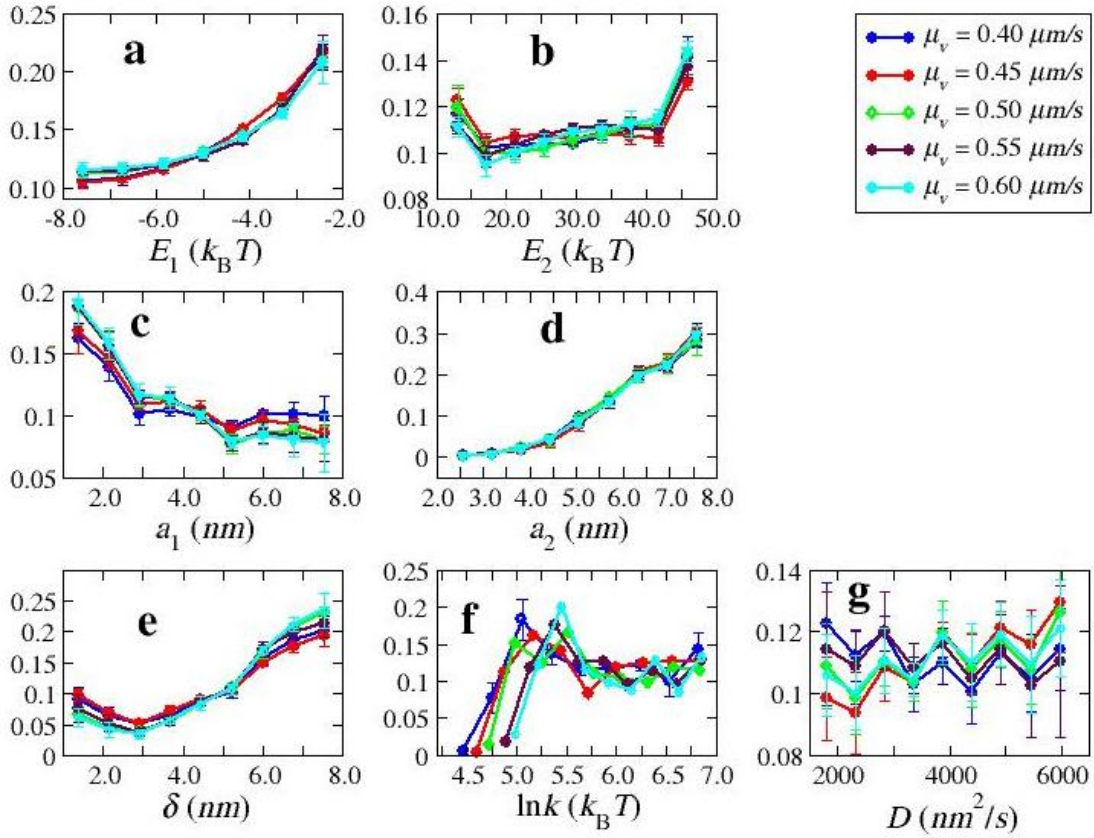

Figure 9. One-variable distributions of $E_1, E_2, a_1, a_2, \delta, \ln k$, and $D$ generated from MC3 simulations. Figures a, b, c, d, e, f, and g show distributions of $E_1, E_2, a_1, a_2, \delta, \ln k$, and $D$, respectively. Blue, red, green, maroon, and cyan symbols denote results corresponding to the input velocity-distributions having average velocities of 0.4, 0.45, 0.50, 0.55, and 0.6 $\mu m/s$, respectively.

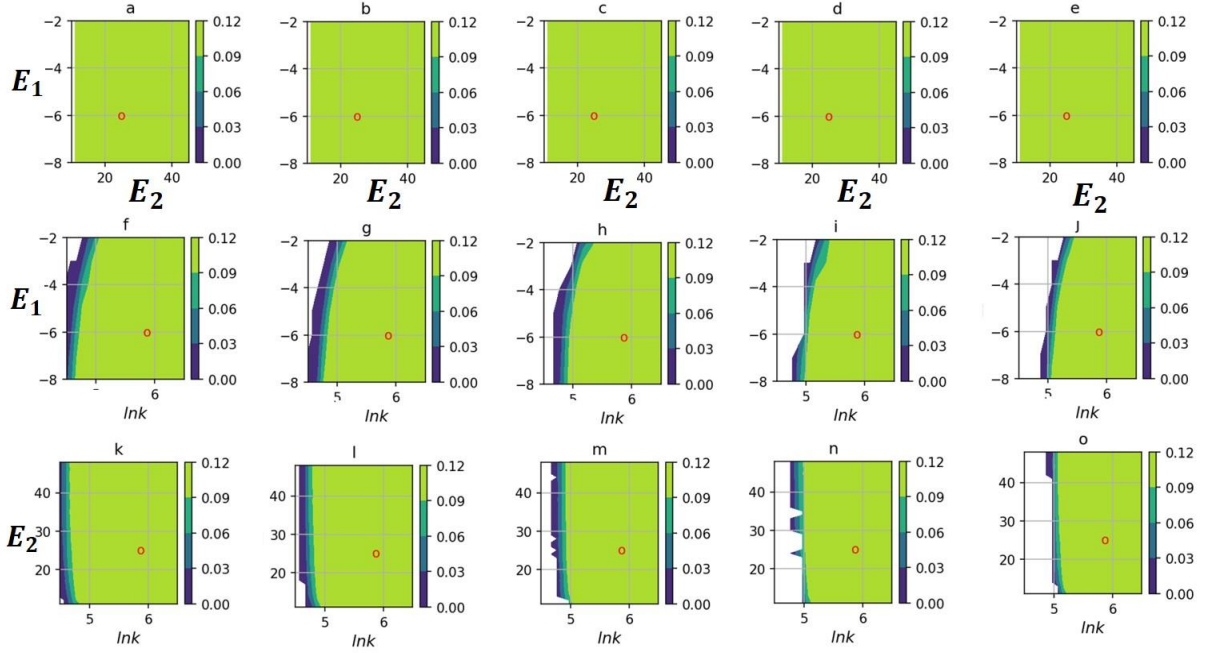

Figure 10. Two-variable distributions obtained from MC3 simulations. Figures a to e show the joint distributions of $E_1$ and $E_2$. Figures f to j show the joint distributions of $E_1$ and $\ln k$. Figures k to o show the joint distributions of $E_2$ and $\ln k$. The figures from left to right in each row denote results corresponding to the input velocity-distributions having average velocities of 0.4, 0.45, 0.50, 0.55, and 0.6 $\mu m/s$, respectively. The red circle in each figure corresponds to $E_1 = -6.0\ k_B T, E_2 = 25.0\ k_B T$, and $\ln k = 5.87\ k_B T$.

### 4.4 Ensemble-averaged potential energy ratchet

The proposed scheme allows us to compute the ensemble-averages of microscopic properties that correspond to the velocity-distribution of the sample. Figure 11 shows the ensemble-averaged $\phi_I(x)$ and $\phi_{II}(x)$ profiles calculated using MC1, MC2, and MC3 simulations. They represent the effective potential energy profiles for the sample of motors. The saw-tooth profiles obtained from MC1 simulations are statistically same for all values of $\mu_v$. The absolute mean values of $E_1$ and $E_2$ are observed to be $5 k_B T$ and $30 k_B T$, respectively. The profiles obtained from MC2 and MC3 simulations are similar. This again indicates the negligible effect of allowing the variation in bulk diffusivity $D$. We also notice curved peaks and valleys due to variable $a_1, a_2$, and $\delta$ in MC2 and MC3 simulations. The average $\phi_I(x)$ profiles flatten slightly with decrease in $\mu_v$, whereas average $\phi_{II}(x)$ profiles show negligible dependence on $\mu_v$. Also, average $\phi_I(x)$ and $\phi_{II}(x)$ profiles obtained from MC2 and MC3 simulations are significantly flatter than those from MC1 simulations. That is, allowing the variation of $a_1, a_2,$ and $\delta$ can help achieve the same velocity-distribution with lower potential energy barrier experienced by motors between stable locations.

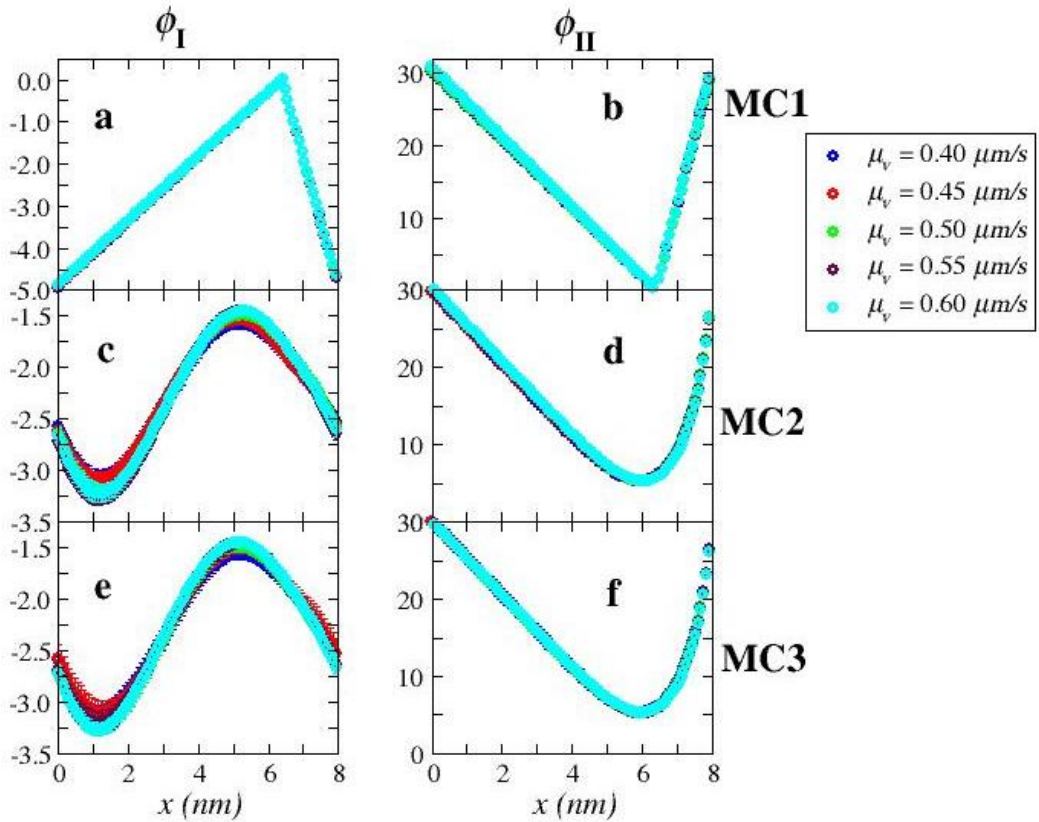

Figure 11. Ensemble-averaged potential energy profiles of states I and II. Figures a, c, and e show the ensemble-averaged potential energy profiles $\phi_I(x)$ of state $I$. Figures b, d, and f show the ensemble-averaged potential energy profiles $\phi_{II}(x)$ of state $II$. Figures a and b show results obtained from MC1 simulations. Figures c and d show results obtained from MC2 simulations. Figures e and f show results obtained from MC3 simulations. Blue, red, green, maroon, and cyan symbols denote results corresponding to the input velocity-distributions having average velocities of 0.4, 0.45, 0.50, 0.55, and 0.6 $\mu m/s$, respectively.

**4.5 Creating a sample of motors having desired mean velocity**

We now summarize the technological relevance of results discussed in Sections 4.1 to 4.4. Let's consider a sample of several nanomotors operating via the fluctuating ratchet mechanism.

The motors have two parts: the body of the motor, and a linear track on which the body translates. The parameters of a fluctuating ratchet model are governed by the chemical features of the body and the track. The parameters $a_1, a_2, L$, and $\delta$ of the model depend on the spatial distribution of chemical species along the track. $E_1$ and $E_2$ depend on the interactions between the body and track, and can be said to denote the maximum binding strength between the track and the body. Thus, a chemical modification of track or body can alter the potential energy profile experienced by the motor in two states. The transitions between states I and II occur via chemical reactions in the body of the motor. The rate-constant $k$ is same for I $\rightarrow$ II and II $\rightarrow$ I reactions, and depends on the chemical features of the body of the motor. The bulk diffusivity $D$ is governed by the size of the body of the motor.

The synthesis of large number of motors will result in some variation in the chemical features of their bodies and tracks, thereby resulting in a distribution of velocities. We assume a Gaussian distribution in velocities with mean $\mu_v$ and standard deviation $\sigma_v$. The goal is to alter the given sample of motors to increase $\mu_v$. Consider the following two scenarios.

**4.5.1 Negligible variation in tracks and sizes of the motor-bodies**

Here, the chemical features of motor-bodies can vary. If the tracks are synthesized with precise control over their chemical features, then $a_1, a_2, L$, and $\delta$ can be assumed to be same for all motors. The bulk diffusivity $D$ will be approximately constant because of negligible variation in the sizes of motor-bodies. $E_1, E_2$, and $\ln k$ can still vary because they depend on the chemical features of the motor-body. Therefore, the above sample corresponds to the one used in MC1 simulations (varying $E_1, E_2, \ln k$). Following the discussion in Section 4.2, increasing the proportion of motors having smaller magnitude of $E_1$ and greater $\ln k$ will increase $\mu_v$. That is, increasing the proportion of motors having weak binding between track and body and faster transitions between two states, increases the mean velocity of the sample.

**4.5.2 Significant variation in tracks and sizes of the motor-bodies**

Due to the variation in chemical features of tracks, the parameters $a_1, a_2, L,$ and $\delta$ will vary along with $E_1$ and $E_2$. The bulk diffusivity $D$ of motors will vary because of the variation in the sizes of motor-bodies. Therefore, this sample corresponds to the one used in MC3 simulations. The discussion in Section 4.4 shows that the average potential energy profile in state I is flatter here than that for precisely controlled tracks and motor-sizes. The discussions in Sections 4.2 and 4.3 also conclude that $\mu_v$ can be increased only by increasing the proportion of motors having greater $\ln k$, or more stable transition state between I and II. Changing the proportion of motors based on the binding between track and body will have negligible effect here.

## 5. Conclusions:

We used Monte Carlo simulations to compute the probability distributions over input parameters of the two-state fluctuating-ratchet model for a given distribution over velocities. The parameters associated with ratchets, transition rates between two states, and motor-sizes were considered. Results were obtained for three cases differentiated by the type of input parameters constrained during Monte Carlo simulations. An analogy with the semi grand canonical ensemble was used to interpret the computed distributions in terms of the proportion of motors having specific features. When all ratchets had same size, teeth-asymmetry, and teeth-locations, then increasing the composition of motors having shorter teeth in one of the states, and faster reaction kinetics increases the mean velocity of the sample. When the motor-size, teeth-asymmetry, and teeth-locations were allowed to vary, then altering the composition based of teeth-heights had negligible effect on the mean velocity. However, increasing the proportion of motors having faster reaction kinetics still increased the mean velocity of the sample. We also calculated the average ratchet of the sample. The average ratchet was flatter when the motor-size, teeth-asymmetry, and teeth-locations were allowed to vary.

The velocity-distributions were assumed to be Gaussian with same standard deviation. The Monte Carlo scheme can be also applied when the given distribution has some another form, and measured over a property other than steady-state velocity. Finally, we note the large statistical uncertainties in the present study due to challenges in sampling the input-parameter space. Flat-histogram based techniques like Wang Landau and transition matrix Monte Carlo can help improve sampling in the future applications.

**SUPPLEMENTARY MATERIAL**

We do not include any supplementary material.

**ACKNOWLEDGEMENTS**

We acknowledge the financial support from the Science and Engineering Research Board (SERB) of India (grants numbered CRG/2020/003225 and MTR/2020/000010) and Department of Science and Technology, Ministry of Science and Technology of India (grant number DST/NSM/R&D_HPC_Applications/2021/2).

**APPENDIX: Comparison with the semi-grand canonical (SG) ensemble**

The probability distributions generated from MC simulations can be interpreted by using the framework of SG ensemble.[9–11] In order to make this comparison, we consider a polydisperse system containing $N$ molecules in a region of volume $V$ and temperature $T$. Note that the molecules are analogous to the motors in our system, and polydispersity is analogous to the variation of microscopic properties of motors and their local environment. We characterize a species in the SG ensemble by using $m$ parameters. Let $\boldsymbol{x}$ be a vector such that the element $x_i$ denotes the value of $i^{th}$ speciation-parameter. We use the variables $\boldsymbol{x}$ and $m$ to be consistent with the description of motors in Section 2.2. The values of input parameters to the fluctuating ratchet model $(x_1, \ldots, x_m)$ can be said to distinguish one species of motor from another. The total number of microstates $\aleph$ in Equation (24) can be said to denote the number of species. If $\aleph \to \infty$, then the partition function of the SG ensemble can be expressed in terms of $N$ integrals over the speciation-parameters:[14]

$$Y = \frac{\exp(-\beta N \mu_r)}{N!} \int \ldots \int Z_N \exp\left(\beta \sum_{i=1}^{N} \mu^*(x^{(i)})\right) \prod_{i=1}^{N} dx^{(i)} \tag{A1}$$

Here, $\beta = 1/k_B T$, $\mu_r$ denotes the chemical potential of reference species. $x^{(i)}$ denotes the vector of speciation-parameters of $i^{th}$ molecule. $\mu^*(x^{(i)}) = \mu(x^{(i)}) + \beta^{-1} \ln q(x^{(i)})$, where $\mu(x^{(i)})$ and $q(x^{(i)})$ denote the chemical potential and internal partition function of species characterized by vector $x^{(i)}$, respectively. $Z_N$ is the configurational partition function given by

$$Z_N = \int \ldots \int \exp(-\beta U(r^N)) \, dr^N \tag{A2}$$

Here, $U(r^N)$ denotes the total energy of system. Note that $U$ depends on the distribution of species in the system. The probability density of observing a species is given by

$$p_{SG}(x) = \frac{1}{N! Y} \int \ldots \int Z_N \exp\left(\beta \sum_{i=1}^{N} (\mu^*(x^{(i)}) - \mu_r)\right) \prod_{i=2}^{N} dx^{(i)} \tag{A3}$$

Here, $p_{SG}(x) dx$ denotes the probability of observing a species having parameters between $x$ and $x + dx$.

The investigated sample of motors can be said to constitute a SG ensemble, where the total number of motors is constant, but the number of motors of each species is allowed to fluctuate. Since a species is characterized by the vector $x$ of microscopic parameters of motor, we can equate $p(x)$ defined in Section 2.2 with $p_{SG}(x)$. For $N \to \infty$, the $l$-variable probability density $\rho^l(x_1, \ldots, x_l)$ can be then said to denote the fraction of motors having particular values of $l$ microscopic parameters. If a microscopic property $x_i$ (or the speciation parameter in SG ensemble) is considered as an order-parameter, then $-\ln \rho^1(x_i)$ is analogous to the potential of mean force (PMF) along $x_i$ in the SG ensemble. The nature of $\rho^1(x_i)$ profile predicted from MC simulations can be then used to infer the relative stability of microscopic states affected by $x_i$. An interesting case is where $\rho^1(x_i)$ is a multimodal distribution, which indicates the coexistence of different microscopic states. That is, the given distribution over an experimental observable may be possible under conditions that result in the coexistence between multiple microscopic states distinguished by $x_i$. On the other hand, a uniform distribution over a particular range of $x_i$ may indicate that the experimental observable is unaffected by the microscopic states distinguished by $x_i$.

If the experimental observable is assumed to be independent of the interaction between motors, then the corresponding SG ensemble can be said to be that of an ideal gas. $p(x)$ can be equated with $p_{SG}^{IG}(x)$, where

$$p_{SG}^{IG}(x) = \frac{\exp[\beta \mu^*(x)]}{\int \exp[\beta \mu^*(x)]\, dx} \tag{A4}$$

Equation (30) can be used to define the chemical potential analogue of a nanoscale motor. If $p(x)$ is derived from the Gaussian distribution over velocities with mean velocity $\mu_v$ and standard deviation $\sigma_v$, then

$$p(x) = \frac{\exp\left[-0.5 \frac{(v(x) - \mu_v)^2}{\sigma_v^2}\right]}{\sigma_v \sqrt{2\pi}} \tag{A5}$$

Comparing Equations (A4) and (A5) gives

$$\mu^*(x) = -0.5 k_B T \frac{(v(x) - \mu_v)^2}{\sigma_v^2} \tag{A6}$$

The chemical potential analogue defined above differs from the real chemical potential of macromolecules that constitute the motor. Nevertheless, $\mu^*(x^{(i)})$ may be useful in extending the framework of SG ensemble to predict the microscopic behaviour of nanoscale motors from the experimental observations.

## AUTHOR DECLARATIONS

**Conflict of interest**

The authors declare no conflicts of interest.

**Author contributions**

**Rupsha Mukherjee:** Literature review, performing simulations, generating plots and graphics.

**Kaustubh Rane:** Writing manuscript, generating plots, interpreting results, acquiring funding.

## DATA AVAILABILITY

The data that support findings will be made available upon request to the corresponding author.